\documentclass[11pt,a4paper]{article}

\usepackage{amssymb}
\usepackage{amsmath}
\usepackage{nicefrac} % for nice fractures
\usepackage{mathtools} % for better ceil brackets, and \coloneqq\coloneqq
\usepackage{enumerate}
\usepackage{latexsym}
\usepackage{float}
\usepackage{color}
\usepackage{xspace}
\usepackage{url}
\usepackage{paralist} %for compact enum etc.
\usepackage{hyperref}
\usepackage{mathdots}
\usepackage{multirow} % multi-cell tables
\usepackage{balance}
\usepackage{subfigure}

\usepackage{tikz}
\usetikzlibrary{arrows, petri,topaths}
\usepackage{tkz-berge}
\usepackage{forest}
\usepackage{listings}
\usepackage{pdfpages}
\usepackage{algorithm}
\usepackage[noend]{algpseudocode}

\usetikzlibrary{automata,positioning}
\tikzset{initial text={}}

\usepackage{tikz}
\usetikzlibrary{arrows, petri,topaths}
\usepackage{tkz-berge}

\usepackage[T1]{fontenc}

\usetikzlibrary{automata,positioning}
\tikzset{initial text={}}

\usetikzlibrary{decorations.markings}

\usepackage{url}
\usepackage{color}
\usepackage{macro}

{\vspace{0.7 em}}

\def\squareforqed{$\blacksquare$}
\def\QED{\ifmmode\squareforqed\else{\unskip\nobreak\hfil
\penalty50\hskip1em\null\nobreak\hfil\squareforqed
\parfillskip = 0pt\finalhyphendemerits = 0\endgraf}\fi}

\renewcommand{\leq}{\leqslant}
\renewcommand{\geq}{\geqslant}

\newcommand\FOtwo{\mbox{$\mathsf{FO}^2$}}

\begin{document}

\title{A simple combinatorial proof for small model property of two-variable logic}

\author{Yanger Ma \qquad\qquad\qquad\qquad Tony Tan
\\
National Taiwan University}

\date{}

\maketitle

\begin{abstract}
We present another proof for the well-known {\em small model property} of two-variable logic.
As far as we know, existing proofs of this property rely heavily on model theoretic concepts.
In contrast, ours is purely combinatorial and uses only a very simple counting argument,
which we find rather intuitive and elegant.
\end{abstract}

%\newpage

\paragraph*{Introduction.}
%\label{sec:intro}

Two-variable logic ($\FOtwo$) is a well known fragment of first-order logic that
uses only two variables and comes with decidable satisfiability problem.
It was first proved to be decidable in double-exponential time by Mortimer~\cite{Mortimer75}.
The upper bound was later improved to single-exponential by Gr\"adel, Kolaitis and Vardi~\cite{GKVardi97}.
Both upper bounds were achieved by showing that $\FOtwo$ has the so called {\em small model property},
i.e., if a formula is satisfiable,
then it is satisfiable by a model with cardinality exponential in the length of the formula,
or double-exponential in the case of Mortimer's.
Both proofs use a rather intricate model theoretic constructions.

In this note we present another proof for the $\FOtwo$ small model property.
The bound we achieve is single-exponential, matching the one by Gr\"adel, et. al.
However, ours is based on a simple graph-theoretic construction from which 
the small model property is a direct implication.
The construction itself is based only on a very simple counting argument.

\paragraph*{A simple graph-theoretic construction.}
\label{sec:lemma}
Recall that a {\em tournament} is a directed graph (with no simple loop) where
there is an edge between every two vertices.  
Let $C,D$ be two disjoint {\em finite} sets of colors
whose elements are called vertex and edge colors, respectively.
A $(C,D)$-graph is a tournament
where the vertices and edges are colored with vertex and edge colors, respectively.
A $(C,D)$-graph is also denoted by $(k,\ell)$-graph,
where $|C|=k$ and $|D|=\ell$.
Note that a $(C,D)$-graph maybe infinite.

In the following we fix a $(C,D)$-graph $G$,
where $|C|=k$ and $|D|=\ell$.
For $c\in C$, let $c(G)$ be the set of vertices in $G$ with color $c$.
We assume that $G$ has {\em fixed orientation} in the sense that between every two vertex colors,
there is only one edge orientation, i.e.,
all edges between vertices in $c_1(G)$ and $c_2(G)$ have the same direction:
either from $c_1$ to $c_2$ or from $c_2$ to $c_1$.

We write $col_G(u,v)$ to denote the color of the edge $(u,v)$ in $G$.
A vertex $u$ in $G$ is incident to an edge color $d$, if there is an edge incident to $u$ with color $d$.
We also say that a vertex $u$ is incident to a pair $(d,c)\in D\times C$,
if there is a vertex $v$ with color $c$ and the edge $(u,v)$ has color $d$.
We write $D_{c_1,c_2}(G)$ to denote the set of the edge colors whose two incident vertices are colored with $c_1$ and $c_2$.

A color $c$ is a {\em king} color (in $G$), if $|c(G)|=1$.
The vertex with a king color is called a {\em king} vertex, or a king, for short.
We denote by $KC(G)$ the set of king colors in $G$.
Obviously, $|KC(G)|$ is precisely the number of kings in $G$.
Let $v_1,\ldots,v_t$ be the kings in $G$ and let
$c_1,\ldots,c_t$ be their respective colors.
For a non-king vertex $u$, the {\em profile of $u$}
is the set $\{(d_1,c_1),\ldots,(d_t,c_t)\}$
where each $d_j$ is the color of the edge connecting $u$ and $v_j$.
Intuitively, the profile of $u$ contains the information about 
the relation between $u$ and each of the kings.

Suppose $k \geq 6$.
We will show that there is a $(k,\ell)$-graph $H$ 
with the same edge orientation between vertex colors as $G$ and the following properties.
\begin{enumerate}[(a)]\itemsep=0pt
\item
$KC(G)=KC(H)$.
\item
For every non-king color $c\in C$, $|c(H)|=k\cdot \ell$.

\item
For every $c_1,c_2 \in C$,
$D_{c_1,c_2}(G)=D_{c_1,c_2}(H)$, i.e.,
the colors of the edges between any two vertices with colors $c_1$ and $c_2$
are the same in both $G$ and $H$.

\item
For every non-king vertex $u$ in $H$,
there is a non-king vertex $v$ in $G$ with the same color and profile as $u$.

\item
For every non-king colors $c_1,c_2$,
for every edge color $d\in D_{c_1,c_2}(G)$,
every vertex $u \in c_1(H)$ is incident to $(d,c_2)$,
i.e., every vertex in $c_1(H)$ is incident to every color in $D_{c_1,c_2}(G)$.
\end{enumerate}

The construction of $H$ is as follows.
For every non-king colors $c$,
we pick pairwise disjoint sets $Z_1^c,\ldots,Z_{k}^c$, where each $|Z_i^c|=\ell$.
We let $Z^c =Z_1^c\cup\cdots \cup Z_k^c$.
The graph $H$ is obtained from $G$ by replacing the vertices in $c(G)$ with $Z^c$,
where all vertices in $Z^c$ are colored with $c$.
The king colors as well as the color and orientation of the edges between king vertices remain the same as in $G$.
Obviously, at this point (a) and (b) already hold $H$.
%We will show how to obtain (c)--(e).

For each non-king color $c$,
we color the edges incident to vertices in $Z^c$ in three steps described below.
The crux of the argument is that since $|c(H)|=k\cdot\ell$,
it is possible to exhaust all the colors so that every vertex in $Z^c$
are incident to every possible edge colors allowed by graph $G$.
In the following we should stress that when coloring the edges,
we keep the orientation between every different two vertex colors the same as in $G$.

%\begin{description}\itemsep=0pt

{\bf (Step~1)}
Coloring the edges between the king vertices and those in $Z^c$.

Recall that $v_1,\ldots,v_t$ are kings with colors $c_1,\ldots,c_t$, respectively.
First, let us pick $t$ sets $Z_1^c,\ldots,Z_t^c$, which is possible since $t\leq k$.
We color the edges between the kings and vertices in $Z_1^c\cup \cdots\cup Z_t^c$
as follows.
For each king $v_i \in \{v_1,\ldots,v_t\}$, we do the following.
\begin{itemize}\itemsep=0pt
\item
We color the edges between $v_i$ and vertices in $Z_i^c$
such that all colors in $D_{c_i,c}(G)$ are used.

\item
Note that we only use the colors in $D_{c_i,c}(G)$ for edges between $v_i$ and $Z_i^c$.
So, for every vertex $u \in Z_i^c$, there is a vertex $x \in c(G)$ such that
$col_H(u,v_i)=col_G(x,v_i)$.
Thus, we can color the edges between $u$ and the rest of the kings (i.e., kings that are not $v_i$)
so that the profile of $u$ in $H$ is the same as the profile of $x$ in $G$.
\end{itemize}
For all the other vertex $u \in Z^c - (Z_1^c\cup\cdots\cup Z_t^c)$,
we pick a vertex $u' \in Z_1^c\cup \cdots \cup Z_t^c$,
and color the edges between $u'$ and the kings so that
both $u'$ and $u$ have the same profile.
At this point (d) is already established.

{\bf (Step~2)} Coloring the edges between vertices in $Z^c$.

Let $s=|D_{c,c}(G)|$.
We re-partition $Z^c$ into three sets $Y_0^c,Y_1^c,Y_2^c$ with each $|Y_i^c|\geq 2\ell$.
For each $i=0,1,2$,
for each vertex $u \in Y_i^c$,
we pick 2s vertices $w_1,\ldots,w_{2s}\in Y_{i+1}^c$.
(When $i=2$, replace $Y_{i+1}^c$ with $Y_0^c$.)
We color the edges $(u,w_1),\ldots,(u,w_s)$ so that
all the colors in $D_{c,c}(G)$ are used.
The orientation of the edges are as indicated, i.e., from $u$ to $w_i$'s.
Similarly, we also color the edges $(w_{s+1},u),\ldots,(w_{2s},u)$
so that all the colors in $D_{c,c}(G)$ are used. 
All the other edges not yet colored
can be colored with arbitrary colors from $D_{c,c}(G)$
and with arbitrary orientation.
Note that in this step, the crux of this step is that $Z^c$ can be re-partitioned
so that all colors in $D_{c,c}(G)$ can be exhausted on every vertex in $Z^c$.
%Note that in this coloring process, 
%we simply exhaust all the colors so
%that every vertex in $Z^c$ are incident to every colors in $D_{c,c}(G)$.

{\bf (Step~3)} Coloring the edges between $Z^{c_0}$ and $Z^c$,
for any non-king color $c_0\neq c$.

Here the coloring process is similar to Step~2,
where we try to exhaust all the colors in $D_{c,c_0}(G)$.
For completeness, we present it here.
Again, let $s=|D_{c,c_0}(G)|$.
We re-partition $Z^c$ into two sets $X_0^c,X_1^c$ with each $|X_i^c|\geq \ell$.
Similarly, $Z^{c_0}$ is re-partitioned into $X_0^{c_0},X_1^{c_0}$ with each $|X_i^{c_0}|\geq \ell$.
The coloring is done as follows.
For each $i=0,1$,
for each vertex $u \in X_i^c$,
we color the edges between $u$ and $X_i^{c_0}$
such that all the colors in $D_{c,c_0}(G)$ are used.
Similarly, for each $i=0,1$,
for each vertex $u \in X_i^{c_0}$,
we color the edges between $u$ and $X_{1-i}^{c}$
such that all the colors in $D_{c,c_0}(G)$ are used.
This marks the end of Step~3.

Similar to Step~2, the crux of the argument in Step~3 is that $Z^c$ and $Z^{c_0}$ can be re-partitioned
so that all colors in $D_{c,c_0}(G)$ can be exhausted on every vertex in $Z^c$ and $Z^{c_0}$.
After Steps~2 and~3, it is straightforward that points (c) and (e) hold.

\paragraph*{Small model property of two-variable logic.}
\label{sec:fo2}

Now we will show how our graph-theoretic construction above can be applied to two-variable logic.
We start with a well known result of Scott~\cite{Scott}
states that every $\fo^2$ sentence  can be converted in polynomial time into the following {\em Scott normal form}:
$$
\Phi \ := \ \forall x \forall y \ \alpha(x,y)  \ \wedge \
\bigwedge_{i=1}^k \forall x \exists y \  \beta_i(x,y)
$$
where $\alpha$ and $\beta_i$ are all quantifier free.
We let $n$ and $m$ to be the number of unary and binary predicates used in $\Phi$.
Adding redundant predicates, if necessary, we assume that $n+m\geq 3$.
We recall a few terminologies.
A {\em $1$-type} is a maximally consistent set of atomic predicates or their negations using only variable $x$.
A {\em $2$-type} is a maximally consistent set of atomic predicates or their negations using variables $x$ and $y$.
Note that both 1-type and 2-type contain atoms such as $r(x,x)$ or its negation, for some binary relation symbol $r$.
The number of $1$-types and $2$-types are $2^{n+m}$ and $2^{2n+4m}$, respectively.

Suppose $\cA\models \Phi$.
The {\em type of an element $a \in A$} is the unique 1-type $\pi$ that $a$ satisfies in $\cA$.
A 1-type is a {\em king}, if there is only one element in $A$ that satisfies it.
Similarly, the type of a pair $(a,b)\in A\times A$ is the unique 2-type that $(a,b)$ satisfies in $\cA$.
We say that a 1-type/2-type being {\em realized} in $\cA$ if there is an element/a pair of elements
that satisfies it.

Note that $\cA$ can be viewed as a complete {\em directed} graph
where the vertex and edge colors are the 1-types and 2-types, respectively.
Since the 2-type of $(a,b)$ uniquely determines the 2-type of its reverse $(b,a)$,
it suffices to consider only one direction.
We fix one direction between every two 1-types realized in $\cA$,
so that $\cA$ can be viewed as $(2^{n+m},2^{2n+4m})$-graph with fixed orientation.
By our construction above,
there is a structure $\cB$ where each non-king 1-type has exactly $2^{3n+5m}$ elements.
Overall, $\cB$ has at most $O(2^{4n+5m})$ elements.
Since both the realized and non-realized types, as well as the profiles of each vertices are preserved in $\cB$,
it is immediate that $\cB\models\Phi$.

\paragraph*{Acknowledgements.}
We acknowledge the generous financial support of Taiwan Ministry of Science and Technology
under grant no. 107-2221-E-002-026-MY2.

\bibliographystyle{abbrv}

\end{document}